\documentclass[12pt]{article}
\usepackage[T2A]{fontenc}
\usepackage[cp1251]{inputenc}
\usepackage{hyperref}
\usepackage{a4wide}
\usepackage{amsmath,latexsym,amsthm,amsfonts}
\usepackage{amssymb,amscd}
\newtheorem{theorem}{Theorem}[section]

\newtheorem{definition}{Definition}
\newtheorem{remark}{Remark}
\newtheorem{lemma}{Lemma}
\newtheorem{example}{Example}

\theoremstyle{remark} 
\numberwithin{equation}{section}

\begin{document}

\title{Sufficient conditions for superstability of many-body interactions}
\author{ M.~V.~Tertychnyi }
\date{}
\maketitle
\begin{footnotesize}
\begin{tabbing}
\= Faculty of physics, Kyiv Shevchenko university , Ukraine \\
\> {\em mt4@ukr.net}\\
\>  \\
\end{tabbing}
\end{footnotesize}
\begin{abstract}
A detailed analysis of necessary conditions on a family of many-body
potentials, which ensure stability, superstability or strong
superstability of a statistical system is given in  present
work.There has been given also an example of  superstable many-body
interaction.
\end{abstract}
\thispagestyle{empty}

\noindent \textbf{Keywords :} Continuous classical system;
many-body interaction; criteria of superstability.\\

\noindent \textbf{Mathematics Subject Classification :}\, 82B05;
 82B21

\section{\bf Introduction}
During last 30 years different sufficient conditions and
restrictions on 2-body potential, which imply superstable or strong
superstable interaction have been studied(see \cite{RT07} for survey
of the results). It is obvious, that  the research of systems with
respect to many-body  interaction requires the same conditions on
potential energy of interaction of any finite quantity of particles
to be fulfilled. In accordance with this fact, one has a similar
problem to describe  the necessary conditions on a sequence of
p-body ($p\geq3)$ potentials, which ensure stability, superstability
or strong superstability of an infinite statistical system. We have
to mention, that such conditions(which ensure an existence of
corellation function in the thermodynamic limit) have been written
in rather abstract form in the works  \cite{Ge82}, \cite{Mo76} and
more implicitly in the works \cite{PR07}, \cite{PS2000},
\cite{ReK04}, \cite{RS97}, \cite{Sk06}, \cite{Za81}. There is
another interesting work in this field(see \cite{BePe2002} ), in
which authors consider a finite sequence of finite range many-body
potentials, one of which is {\it stabilizing}, and ensures stability
of a whole system. In the present paper we consider an infinite
system of finite range many-body potentials  taking into account the
traditional concept, i.e. in some sense p-body potential plays less
important role in the total energy of interaction than $p-1$-one.
Each of p-body potentials can be both positive or negative and it
depends on the configuration of particles. The conditions on a
sequence of p-body ($p> 2$) potentials, which ensure stability,
superstability or strong superstability of a system, if such a
behavior is enabled by 2-body(pair) potential of interaction are
formulated in this article. In the next section we give necessary
definitions and formulate main result. In section~4 we give  an
example of many-body interaction, which yields above mentioned
conditions
\section{\bf Definitions and main result}

Let $\mathbb{R}^{d}$ be a $\textnormal{d}$-dimensional Euclidean
space. Following [9] for each $r\in \mathbb{Z}^{d}$ and
$\lambda\in\mathbb{R}_{+}$ we define an elementary cube with a rib
$\lambda$ and center $r$:

\begin{equation}\label{E:part}
\Delta_{\lambda}(r)=\left\{ x\in \mathbb{R}^{d}\; \mid
\;\lambda\left(r^{i}-1/2 \right)\leq x^{i}< \lambda \left( r^{i}+1/2
\right)\right\}
\end{equation}
We will sometimes write $\Delta$ instead of $\Delta_{\lambda}(r)$,
if a cube $\Delta$ is considered to be arbitrary and there is no
reason to emphasize that it is centered in the particular point
$r\in\mathbb{Z}^{d}$. We denote by $\overline{\Delta_{\lambda}}$
the corresponding partition of $\mathbb{R}^{d}$ into cubes
$\Delta$.
 Let us consider a general type of many-body interaction specified by a
family of p-body potentials\,
$V_{p}:\left(\mathbb{R}^{d}\right)^{p}\rightarrow\mathbb{R},\,
p\geq2 $ and define also positive and negative parts  of interaction
potential:
\[
V_{p}^{+}\left(x_{1},\ldots,x_{p}\right):=
\textnormal{max}\left\{0;\;
V_{p}\left(x_{1},\ldots,x_{p}\right)\right\},
\]
\[
V_{p}^{-}\left(x_{1},\ldots,x_{p}\right):=
\textnormal{min}\left\{0;\;
V_{p}\left(x_{1},\ldots,x_{p}\right)\right\}
\]
We assume for the family of potentials
$V:=\,\left\{V_{p}\right\}_{p\geq2}$ the following conditions:

{\bf A1. Symmetry.}\; For any $p\geq2$, any
$(x_{1},\ldots,x_{p})\in\left(\mathbb{R}^{d}\right)^{p}$ and any
permutation $\pi$ of the numbers $\{1,\ldots, p\}$:

\[V_{p}\left(x_{1},\ldots,x_{p}\right)=V_{p}\left(x_{\pi(1)},\ldots,x_{\pi(p)}\right).\]

{\bf A2. Translation invariance.}\; For any $p\geq2$, any
$\left(x_{1},\ldots,x_{p}\right)\in\left(\mathbb{R}^{d}\right)^{p}$
and $a\in \mathbb{R}^{d}$:

\[V_{p}\left(x_{1},\ldots,x_{p}\right)=V_{p}\left(x_{1}+a,\ldots,x_{p}+a\right).\]

{\bf A3. Repulsion for small distances.}\; There exists a
partition of $\mathbb{R}^{d}$ into cubes
$\overline{\Delta_{\lambda}}$ \;(see~\eqref{E:part}) such that for
any $\left(x_{1},\ldots,x_{p}\right)\subset\Delta,\; p\geq2$:\,
$V_{p}\left(x_{1},\ldots,x_{p}\right)\geq 0$.

{\bf A4. Integrability.}\;
\begin{equation}\label{E:integr}
\underset{\left\{x_{1},\ldots,x_{k}\right \}\in
(\mathbb{R}^{d})^{k}}{\text{sup}}\quad
\underset{\left(\mathbb{R}^{d}\right)^{p-k}} \int \; \left|
V_{p}^{-} \,\left(x_{1},\ldots,x_{p}\right )
\right|\;dx_{k+1}\cdot\ldots\cdot dx_{p}<+\infty\, , \quad 1\leq
k\leq p-1.
\end{equation}

Under assumptions {\bf A1-A4} we introduce the energy
$U\left(\gamma \right):\quad \Gamma_{0}\rightarrow
\mathbb{R}\cup{\{+\infty\}}$, which corresponds to the family of
potentials
$V_{p}:\left(\mathbb{R}^{d}\right)^{p}\rightarrow\mathbb{R},\,
p\geq2 $ and which is defined by:
\begin{equation}\label{E:energy}
U(\gamma)=\sum_{p\geq2}\; \sum_{\{x_{1},\ldots,x_{p}\}\subset{
\gamma }}V_{p}(x_{1},\ldots,x_{p})\, , \gamma\in \Gamma_{0},\;
\left|\gamma \right |\geq2,
\end{equation}
where $\Gamma_{0}$ is the space of finite configurations
\begin{equation}\label{E:conf}
  \Gamma_{0}= \underset{n\in \mathbb{N}_{0}}{\coprod}\Gamma^{(n)}\,
  , \quad \Gamma^{(n)}:=\left \{\gamma \subset \mathbb{R}^{d} \; | \;
  |\gamma|=n \right \},\; \mathbb{N}_{0}=\mathbb{N}\cup\{0\},\;
  \Gamma^{(0)}=\emptyset.
\end{equation}
Let's  consider also the part of a total energy, defined only by
p-body potential:
\begin{equation}\label{E:penerg}
U^{(p)}(\gamma)= \underset{{\{x_{1},\ldots,x_{p}\}\subset{ \gamma
}}}{\sum\;}V_{p}(x_{1},\ldots,x_{p}),\;\gamma\in \Gamma_{0},\;
|\gamma|\geq2.
\end{equation}

We introduce 3 kinds of interactions, defined by the family of
potentials $V:=\,\left\{V_{p}\right\}_{p\geq2}$.
\begin{definition}\label{D: stab}
Interaction, defined by the family of potentials
$V:=\,\left\{V_{p}\right\}_{p\geq2}$\,is called:\\
a)\;stable, if there exists $B$>0 such that:\\
\begin{equation}\label{E:stable}
U(\gamma)\geq-B|\gamma|,\; \text{for any\;}\gamma \in \Gamma_{0};
\end{equation}
b)\;superstable, if there exist $A>0, \,B\geq0 $
and partition into cubes $\overline{\Delta_{\lambda}}$ such that:\\
\begin{equation}\label{E:sstable}
U(\gamma)\geq A \underset{\Delta\in \overline{\Delta_{\lambda}} }
{\sum}\; |\gamma_{\Delta}|^2 - B|\gamma|,\; \text{for any\;}\gamma
\in \Gamma_{0};
\end{equation}
c)\;strong superstable, if there exist $A>0,
\,B\geq 0,\;m>2\;$ and partition into cubes $\overline{\Delta_{\lambda}}$ such that:\\
\begin{equation}\label{E:ssststable}
U(\gamma)\geq A \underset{\Delta\in \overline{\Delta_{\lambda}} }
{\sum}\; |\gamma_{\Delta}|^m - B|\gamma|,\; \text{for any\;}\gamma
\in \Gamma_{0}.
\end{equation}
\end{definition}
In the above conditions   constants  $A,B$  can depend on
$\overline{\Delta_{\lambda}}$ and consequently on $\lambda$. In our
future estimates we will use several notations, which we introduce
below.
\begin{definition}\label{D:sum}
Let $\Delta,\,\Delta_{i}\in
\overline{\Delta_{\lambda}};\,n,m,k_{i},k \in \mathbb{N},
k_{1}+\cdots+k_{n}=p,\,p\geq 2$. Then:
\begin{align}
&a)\;I_{p}^{k_{1},\ldots,k_{n}}(\Delta_{1},\ldots,\Delta_{n}):=\underset{\left\{x_{1}^{(1)},\ldots,x_{k_{1}}^{(1)}\right\}\subset\Delta_{1},
\ldots,\left\{x_{1}^{(n)},\ldots,x_{k_{n}}^{(n)}\right\}\subset\Delta_{n}}{\textnormal{sup}}\;
\left|V_{p}^{-}\left(x_{1}^{(1)},\ldots,x_{k_{n}}^{(n)}\right)\right|,\label{E:sum1}\\
&b)\;I_{p}^{k|m}(\Delta):=\underset{(
\Delta_{1},\ldots,\Delta_{m})\subset\overline{\Delta_{\lambda}}}{\sum}\,I_{p}^{k,\overbrace{1,\ldots,1}^{m}}(\Delta,\Delta_{1},\ldots,\Delta_{m}),\;
k+m=p. \label{E:sum2}
\end{align}
The sum in \eqref{E:sum2} means independent sums w.r.t. every
$\Delta_{i},\, i=\overline{1,m}$.
\end{definition}
\begin{definition}\label{D:sum1}
Under the conditions of the  def.~2 let
\;$\Delta_{i}\neq \Delta_{j},\text{if}\;i\neq j,\\
\Delta_{i}\neq\Delta,\; \;1\leq i\leq m$. This means that all
cubes are different. Then:
\begin{equation}\label{E:sum3}                                              
I_{p}^{k| \{
k_{1},\ldots,k_{m}\}}(\Delta):=\underset{\{\Delta_{1},\ldots,\Delta_{m}\}\subset\overline{\Delta_{\lambda}}}{\sum}\;
\sum^{\prime}_{\pi\in
P_{m}}\,I_{p}^{k,\,k_{\pi(1)},\ldots,k_{\pi(m)}}(\Delta,\Delta_{1},\ldots,\Delta_{m}),
\end{equation}
where $P_{m}$ is  a set of all permutations of
numbers\,\{1,\ldots,m\}, but the sum $\sum^{\prime}_{\pi\in P_{m}}$
takes into account only different permutations of numbers
$\{k_{1},\ldots,k_{m}\}$ (for example if $k_{i}=k_{j}$ for some
$i,j$, permutation of numbers $k_{i}, k_{j}$ is considered only
once).
\end{definition}
There are three useful remarks and two lemmas, which are closely
connected with sums\,\eqref {E:sum2}, \eqref{E:sum3}
\begin{remark}\label{R:subst}
From the above definitions the following inequality holds:
\[I_{p}^{k| \{
k_{1},\ldots,k_{i},\ldots,k_{m}\}}(\Delta)=I_{p}^{k_{i}| \{
k_{1},\ldots,k,\ldots,k_{m}\}}(\Delta).\]
\end{remark}
\begin{remark}\label{R:converg}
If $\lambda\rightarrow 0$:
\begin{equation}\label{E:converg}
\lambda^{md}I_{p}^{k|m}(\Delta)\rightarrow
\underset{\{x_{1},\cdots,x_{k}\}\subset\Delta}{\textnormal{sup}}\,\int_{R^m}
\left|V_{p}^{-}(x_{1},\ldots,x_{k},x_{k+1}\ldots,x_{k+m})\right|\,dx_{k+1}\cdot\ldots\cdot\,dx_{k+m}.
\end{equation}
It allows us to write an estimate for the value of
$I_{p}^{k|m}(\Delta)$.
\end{remark}
\begin{remark}\label{R:transl_inv}
 Due to the assumption {\bf A2} value of $I_{p}^{k|m}(\Delta)$
 does not depend on the position of cubes $\Delta$, so we can put
\begin{equation}\label{E:transl_inv}
I_{p}^{k|m}(\Delta)=I_{p}^{k|m}.
\end{equation}
\end{remark}

\begin{lemma}\label{L:sumint}
For any $p\geq2$ the following inequality holds:
\begin{equation}\label{E:sumint}
\sum_{j=2}^{p}\,\sum_{\substack{k_{l}\geq 1,\; 1\leq l \leq j,\\
k_{1}+\cdots+k_{j}=p,\\k_{1}\leq\ldots\leq k_{j} }} \,
I_{p}^{k_{1}\,|\,\{k_{2},\ldots,k_{j}\}}(\Delta)\leq
I_{p}^{1\,|\,p-1}(\Delta).
\end{equation}
\end{lemma}
\emph{Proof.} Using the definition \eqref{E:sum2} we can rewrite
$I_{p}^{1\,|\,p-1}(\Delta)$ in the following form:
\begin{equation}\label{E:lemmaproof1}
I_{p}^{1\,|\,p-1}(\Delta)=\underset{(
\Delta_{2},\ldots,\Delta_{p})\subset\overline{\Delta_{\lambda}}}{\sum}\,I_{p}^{\overbrace{1,\ldots,1}^{p}}(\Delta,\Delta_{2},\ldots,\Delta_{p})
\
\end{equation}
The sum in the r.h.s of \eqref{E:lemmaproof1} can be rewritten in
the form of sums over sets of nonintersecting cubes
$\{\Delta_{2},\ldots, \Delta_{j}\}, j=\overline{2,p}$, which belong
to the area $\overline{\Delta_{\lambda}}\setminus \{\Delta\}$. Then,
neglecting some combinatoric coefficients, which are greater then
unity and as $I_{p}^{p}(\Delta)\equiv 0$ for sufficiently small
$\lambda$ (see {\bf A3} and Eq.~\eqref{E:sum1}),
equation~\eqref{E:lemmaproof1} can be represented in the form of
inequality:
\begin{equation}\label{E:lemmaproof2}                                                                 
I_{p}^{1\,|\,p-1}(\Delta) \geq \sum_{j=2}^{p}\,\sum_{\substack{k_{l}\geq 1,\; 1\leq l \leq j,\\
k_{1}+\cdots+k_{j}=p,\\k_{1}\leq\ldots\leq k_{j} }}
\,\sum_{\{\Delta_{2},\ldots,\Delta_{j}\}\subset\overline{\Delta_{\lambda}}\setminus
\{\Delta\}}\; \sum^{\prime}_{\pi\in P_{j}
}\,I_{p}^{k_{\pi(1)},\ldots,k_{\pi(j)}}(\Delta,\Delta_{2},\ldots,\Delta_{j})
\end{equation}
Let us take into account the following obvious estimate:
\begin{equation}\label{E:lemmaproof3}                                                             
\sum^{\prime}_{\pi\in P_{j}
}\,I_{p}^{k_{\pi(1)},\ldots,k_{\pi(j)}}(\Delta,\Delta_{2},\ldots,\Delta_{j})\geq
\sum^{\prime}_{\pi \in  P_{j \backslash \{1\}}
}\,I_{p}^{k_{1},k_{\pi(2)}\ldots,k_{\pi(j)}}(\Delta,\Delta_{2},\ldots,\Delta_{j}),
\end{equation}
where $P_{j\backslash\{1\}}$ is a set of all permutations of
numbers $\{2,\ldots,j\}$.
 Using \eqref{E:sum3},\,\eqref{E:lemmaproof2},\,
\eqref{E:lemmaproof3}, we obtain finally:
\begin{equation}\label{E:lemmaproof4}
I_{p}^{1\,|\,p-1}(\Delta)\geq
\sum_{j=2}^{p}\,\sum_{\substack{k_{l}\geq 1,\; 1\leq l \leq j,\\
k_{1}+\cdots+k_{j}=p,\\k_{1}\leq\ldots\leq k_{j} }} \,
I_{p}^{k_{1}\,|\,\{k_{2},\ldots,k_{j}\}}(\Delta). \notag
\end{equation}
$$
\mspace{675mu}\blacksquare
$$
\begin{lemma}\label{L:sumint2}
For any $p\geq 2$ the following inequality holds:
\begin{equation}\label{E:inequal1}                                                        
\sum_{\substack{\{ \Delta_{1},\ldots,\Delta_{j}\}\subset
\overline{\Delta_{\lambda}},\\
|\gamma_{\Delta_{r}}|\geq 1,\, 1\leq r\leq j}}\,
\sum^{\prime}_{\pi\in P_{j}
}\,I_{p}^{k_{\pi(1)},\ldots,k_{\pi(j)}}(\Delta_{1},\ldots,\Delta_{j})
\sum_{i=1}^{j}|\gamma_{\Delta_{i}}|^{p}\leq
 j \sum_{\substack{\Delta \in \overline{\Delta_{\lambda}},
\\|\gamma_{\Delta}|\geq 1}}
|\gamma_{\Delta}|^{p}I_{p}^{k_{1}|\{k_{2},\ldots,k_{j}\}}(\Delta).
\end{equation}
\end{lemma}
\emph{Proof.} \,From the  def.~3 (see \eqref{E:sum3}) and taking
into account, that if we split the first sum in \eqref{E:inequal1}
into $j$ independent sums over\, $\Delta_{i}\in
\overline{\Delta_{\lambda}},\, i= \overline{1,j};\, \Delta_{l}\neq
\Delta_{k},\textnormal{if}\, l\neq k$\, the number of terms
increases in\, $j!$\, times, so
\begin{align}\label{L2:est1}                                                                
&L:=\sum_{\substack{\{ \Delta_{1},\ldots,\Delta_{j}\}\subset
\overline{\Delta_{\lambda}},\\
|\gamma_{\Delta_{r}}|\geq 1,\, 1\leq r\leq j}}\,
\sum^{\prime}_{\pi\in P_{j}
}\,I_{p}^{k_{\pi(1)},\ldots,k_{\pi(j)}}(\Delta_{1},\ldots,\Delta_{j})
\sum_{i=1}^{j}|\gamma_{\Delta_{i}}|^{p}=\notag \\
&=\frac{1}{j!}\sum_{\substack{
\Delta_{1}\in\overline{\Delta_{\lambda}},\ldots,
\Delta_{j}\in\overline{\Delta_{\lambda}}\\\Delta_{l}\neq
\Delta_{k},\, l\neq k}}\;\;\sum^{\prime}_{\pi\in P_{j}
}\,I_{p}^{k_{\pi(1)},\ldots,k_{\pi(j)}}(\Delta_{1},\ldots,\Delta_{j})\sum_{i=1}^{j}|\gamma_{\Delta_{i}}|^{p}.
\end{align}

For any $\{\Delta_{1},\ldots,\Delta_{j}\}\subset
\overline{\Delta_{\lambda}}$ the following estimate is true:
\begin{equation}\label{L2:est2}                                                              
\sum^{\prime}_{\pi\in P_{j}
}\,I_{p}^{k_{\pi(1)},\ldots,k_{\pi(j)}}(\Delta_{1},\ldots,\Delta_{j})\leq
\sum_{t=1}^{j}\sum^{\prime}_{\pi\in P_{j} \setminus\{t\}
}\,I_{p}^{k_{t},k_{\pi(2)},\ldots,k_{\pi(j)}}(\Delta_{1},\ldots,\Delta_{j}).
\end{equation}
We obtain from \eqref{L2:est1}, \eqref{L2:est2}:
\begin{equation}\label{L2:est3}                                                                     
 L\leq \frac{1}{j!}\sum_{r=1}^{j}\;\;\sum_{\Delta_{r}\in\overline{\Delta_{\lambda}}}\sum_{\substack{
\Delta_{1},\ldots,\Delta_{r-1}\in\overline{\Delta_{\lambda}},\\
\Delta_{r+1},\ldots, \Delta_{j}\in\overline{\Delta_{\lambda}},\\
\Delta_{l}\neq \Delta_{k},\, l\neq k}}\; \sum_{t=1}^{j}\;
\sum^{\prime}_{\pi\in P_{j} \setminus\{t\}
}\,I_{p}^{k_{t},k_{\pi(2)},\ldots,k_{\pi(j)}}(\Delta_{1},\ldots,\Delta_{j})|\gamma_{\Delta_{r}}|^{p}.
\end{equation}
As the number of sets
$\{\Delta_{1},\ldots,\Delta_{r-1},\Delta_{r+1},\ldots,\Delta_{j}\}\subset
\overline{\Delta_{\lambda}} $ in the third group of sums in
\eqref{L2:est3} is $(j-1)!$ and taking into account the def.~3
(see\eqref{E:sum3}) one can rewrite \eqref{L2:est3} in the following
way:
\begin{equation}\label{L2:est4}
L\leq
\frac{1}{j}\sum_{r=1}^{j}\;\sum_{\Delta_{r}\in\overline{\Delta_{\lambda}}}\;\sum_{t=1}^{j}
\,I_{p}^{k_{t}|\{k_{1},\ldots,k_{t-1},k_{t+1},\ldots,k_{j}\}}(\Delta_{r})\,
|\gamma_{\Delta_{r}}|^{p}.
\end{equation}
We deduce finally from the  remarks~1,3  and \eqref{L2:est4}:
\begin{equation*}
L\leq  j\sum_{\substack{\Delta \in \overline{\Delta_{\lambda}},
\\|\gamma_{\Delta}|\geq 1}}
|\gamma_{\Delta}|^{p}I_{p}^{k_{1}|\{k_{2},\ldots,k_{j}\}}(\Delta).
\end{equation*}
$$
\mspace{675mu}\blacksquare
$$
We give the following definition for the positive part of
interaction potential:
\begin{equation}\label{E:pospart}
V_{p}^{p}(\Delta):=\underset{\{x_{1},\ldots,x_{p}\}\subset \Delta
}{\textnormal{inf}}V_{p}^{+}(x_{1},\ldots,x_{p})
\end{equation}

The main result of the article is in the following theorem:
\begin{theorem}\label{T:main}
Let the family of $\textnormal{p}$-body potentials
$V_{p}:\left(\mathbb{R}^{d}\right)^{p}\rightarrow\mathbb{R},\,
p\geq2$ satisfy \\ assumptions {\bf A1-A4}. Let also the part of
interaction $U^{(2)}(\gamma)$ be stable (superstable, strong
superstable). If there exists such partition of $\mathbb{R}^{d}$
into cubes $\overline{\Delta_{\lambda}}$, that for each $p>2$
\\the following holds:
\begin{align}
&1) \frac{V_{p}^{p}(\Delta)}{p^p}-p\,I_{p}^{1|p-1}(\Delta)\geq
0; \label{E:main}\\
&2)
\sum_{p>2}p^{p+1}I_{p}^{1|p-1}(\Delta)<+\infty.\label{E:main_addit}
\end{align}
then interaction, corresponding to this family of potentials, is
also stable (superstable, strong superstable).
\end{theorem}
\section{\bf  Proof of Theorem \ref{T:main} }

\emph{Proof.} Let conditions of the theorem \eqref{T:main} hold
and  $\gamma\in \Gamma_{0}$. We can write $U^{(p)}(\gamma)$ in the
following form:
\begin{align}\label{E:est1}
  &U^{(p)}(\gamma)=
  \sum_{\substack{\Delta\in \overline{\Delta_{\lambda}},\\ |\gamma_{\Delta}|\geq
  p}}\,\sum_{\{x_{1},\ldots,x_{p}\}\subset\gamma_{\Delta}}V_{p}(x_{1},\ldots,x_{p})+ \notag\\
  &+\sum_{j=2}^{p}\,\sum_{\substack{k_{l}\geq1,\,1\leq l\leq j,\\k_{1}+\cdots+k_{j}=p,\\k_{1}\leq\ldots\leq
  k_{j}}}\,
  \sum_{\substack{\{ \Delta_{1},\ldots,\Delta_{j}\}\subset
  \overline{\Delta_{\lambda}},\\
  |\gamma_{\Delta_{r}}|\geq 1,\, 1\leq r\leq j}}\,\sum^{\prime}_{\substack{\pi:\,k_{_{\pi(n)}}\leq |\gamma_{\Delta_{n}}|,\\
  1\leq n\leq j}}\times\\
  & \times
  \sum_{
  \{x_{1}^{(1)},\ldots,x_{k_{\pi(1)}}^{(1)}\}\subset \gamma_{\Delta_{1}},\ldots,
  \{x_{1}^{(j)},\ldots,x_{k_{\pi(j)}}^{(j)}\}\subset
  \gamma_{\Delta_{j}}}
  V_{p}(
  x_{1}^{(1)},\ldots, x_{k_{\pi(j)}}^{(j)}).\notag
\end{align}
The first part of  \eqref{E:est1} includes the interaction of
particles within every arbitrary cube $\Delta$, the second one does
the same with particles, which are situated in different cubes of
$\overline{\Delta_{\lambda}}$  with $|\gamma_{\Delta}|\geq 1$. The
4-th group of sums in the second term of \eqref{E:est1} is the sum
over all  different permutations (see def.~3) $\pi: (k_{1},\ldots,
k_{j})\rightarrow (k_{\pi(1)},\ldots, k_{\pi(j)})$ and all values
$k_{1},\ldots, k_{j}(k_{1}\leq \ldots \leq k_{j}),\; k_{l}\geq 1, l=
\overline{1,j}, \, k_{1}+\cdots+k_{j}=p$ with the restrictions
$1\leq k_{\pi(n)}\leq |\gamma_{\Delta_{n}}|, n=\overline{1,j}$.
 Let us explain this notation by simple
example. Let the number of cubes, where there are particles for
7-potential be $j=4$. The set of $k_{i}$ is $(1,\,2,\,2,\,2)$.\,We
consider a set of cubes $\{\Delta_{1},\ldots,\Delta_{4}\}$ such that
\,$|\gamma_{\Delta_{1}}|=1,\,|\gamma_{\Delta_{2}}|=3,\,|\gamma_{\Delta_{3}}|=2,\,|\gamma_{\Delta_{4}}|=6.\,$
As  a result, all permutations $\pi$ such that
$\pi(1)=2,\,\pi(1)=3,\,\pi(1)=4$ are not allowed, i.e.
$k_{2}=k_{3}=k_{4}=2>|\gamma_{\Delta_{1}}|$.\, Using definitions
\eqref{E:pospart} and \eqref{E:sum1} we can estimate
\eqref{E:est1}\,in the following way:
\begin{align}\label{E:est2}
&U^{(p)}(\gamma)\geq \sum_{\substack{\Delta\in
\overline{\Delta_{\lambda}},\\|\gamma_{\Delta}|\geq
p}}V_{p}^{p}(\Delta)C_{|\gamma_{\Delta}|}^{p}-
\sum_{j=2}^{p}\,\sum_{\substack{k_{l}\geq1,\,1\leq l\leq
j,\\k_{1}+\cdots+k_{j}=p,\\k_{1}\leq\ldots\leq k_{j}}}\,
\sum_{\substack{\{ \Delta_{1},\ldots,\Delta_{j}\}\subset
\overline{\Delta_{\lambda}},\\
|\gamma_{\Delta_{r}}|\geq 1,\, 1\leq r\leq
j}}\,\sum^{\prime}_{\substack{\pi:\,k_{_{\pi(n)}}\leq
|\gamma_{\Delta_{n}}|,\\
1\leq n\leq j}}\times \\
&\times \left(\prod_{m=1}^{j}
C_{|\gamma_{\Delta_{m}}|}^{k_{\pi(m)}}\right)\cdot\;I_{p}^{k_{\pi(1)},\ldots,k_{\pi(j)}}(\Delta_{1},\ldots,\Delta_{j}),\notag
\end{align}
where $C_{n}^{k}=\frac{n!}{(n-k)!k!}$.  Using inequalities: $
\forall n\geq k \geq 1,\, \frac{n^k}{k^k}\leq C_{n}^{k}\leq
\frac{n^k}{k!} $, we obtain:
\begin{align}\label{E:est3}
&U^{(p)}(\gamma)\geq \sum_{\substack{\Delta\in
\overline{\Delta_{\lambda}},\\|\gamma_{\Delta}|\geq p}}\frac
{V_{p}^{p}(\Delta)}{p^p}|\gamma_{\Delta}|^{p}-
\sum_{j=2}^{p}\,\sum_{\substack{k_{l}\geq1,\,1\leq l\leq
j,\\k_{1}+\cdots+k_{j}=p,\\k_{1}\leq\ldots\leq k_{j}}}\,
\sum_{\substack{\{ \Delta_{1},\ldots,\Delta_{j}\}\subset
\overline{\Delta_{\lambda}},\\
|\gamma_{\Delta_{r}}|\geq 1,\, 1\leq r\leq j}}\times \\
&\times \sum^{\prime}_{\substack{\pi:\,k_{_{\pi(n)}}\leq
|\gamma_{\Delta_{n}}| ,\\
1\leq n\leq j}}
I_{p}^{k_{\pi(1)},\ldots,k_{\pi(j)}}(\Delta_{1},\ldots,\Delta_{j})
 \prod_{m=1}^{j}
\frac{|\gamma_{\Delta_{m}}|^{k_{\pi(m)}}}{k_{\pi(m)}!}.\notag
\end{align}
Let us consider the following inequality (see   Appendix for the
proof ):
\begin{equation}\label{E:estappendix}
\prod_{i=1}^{j}a_{i}^{m_{i}}\leq\frac{1}{m_{1}+\cdots+m_{j}}\sum_{i=1}^{j}m_{i}\,a_{i}^{m_{1}+\cdots+m_{j}}\leq
\sum_{i=1}^{j}\,a_{i}^{m_{1}+\cdots+m_{j}},
\end{equation}
where $a_{1},\dots,a_{j}\in \mathbb{R}_{+};m_{1},\dots,m_{j}\in
\mathbb{N}$.
 Using \eqref{E:estappendix}, we obtain:
\begin{align}\label{E:est4}
&U^{(p)}(\gamma)\geq \sum_{\substack{\Delta\in
\overline{\Delta_{\lambda}},\\|\gamma_{\Delta}|\geq p}}\frac
{V_{p}^{p}(\Delta)}{p^p}|\gamma_{\Delta}|^{p}-
\sum_{j=2}^{p}\,\,\sum_{\substack{k_{l}\geq1,\,1\leq l\leq
j,\\k_{1}+\cdots+k_{j}=p,\\k_{1}\leq\ldots\leq k_{j}}}\,
\sum_{\substack{\{ \Delta_{1},\ldots,\Delta_{j}\}\subset
\overline{\Delta_{\lambda}},\\
|\gamma_{\Delta_{r}}|\geq 1,\, 1\leq r\leq j}}\times\\
&\times
\prod_{m=1}^{j}\frac{1}{k_{m}!}\sum^{\prime}_{\substack{\pi:\,k_{_{\pi(n)}}\leq
|\gamma_{\Delta_{n}}| ,\\
1\leq n\leq
j}}\,I_{p}^{k_{\pi(1)},\ldots,k_{\pi(j)}}(\Delta_{1},\ldots,\Delta_{j})
\sum_{i=1}^{j}|\gamma_{\Delta_{i}}|^{p}.\notag
\end{align}
Taking into account, that the sum w.r.t. $\pi$ defined in
\eqref{E:est1} contains less number of terms, the the same one in
\eqref{E:sum3}, as it does not have the restrictions
$k_{\pi(n)}\leq|\gamma_{\Delta_{n}}|,\, n=\overline{1,j}$ and using
 lemma~2, that is inequality \eqref{E:inequal1}, we obtain:
\begin{equation}\label{E:est6}
U^{(p)}(\gamma)\geq \sum_{\substack{\Delta\in
\overline{\Delta_{\lambda}},\\|\gamma_{\Delta}|\geq p}}\frac
{V_{p}^{p}(\Delta)}{p^p}|\gamma_{\Delta}|^{p}-
\sum_{j=2}^{p}\,\frac{j}{B(p;j)}\,\sum_{\substack{\Delta\in
\overline{\Delta_{\lambda}},\\|\gamma_{\Delta}|\geq
1}}|\gamma_{\Delta}|^{p} \,\sum_{\substack{k_{l}\geq1,\,1\leq l\leq
j,\\k_{1}+\cdots+k_{j}=p,\\k_{1}\leq\ldots\leq
k_{j}}}I_{p}^{k_{1}|\{k_{2},\ldots,k_{j}\}}(\Delta),
\end{equation}
where $$B(p;j)=\underset{\substack{k_{\pi(t)}\geq 1,\,1\leq t\leq
p,\\k_{\pi(1)}+\cdots+k_{\pi(j)}=p}}{\textnormal{inf}}(k_{\pi(1)}!\cdot\ldots\cdot
k_{\pi(j)}!)$$ Since\;$\textnormal{max}\frac{j}{B(p;j)}=p,\, 2\leq
j\leq p$ and taking into account  definitions~2, 3 and  lemma~1, we
deduce, that:
\begin{equation}\label{E:est7}
U^{(p)}(\gamma)\geq \sum_{\substack{\Delta\in
\overline{\Delta_{\lambda}},\\|\gamma_{\Delta}|\geq p}}\frac
{V_{p}^{p}(\Delta)}{p^p}|\gamma_{\Delta}|^{p}-p\,I_{p}^{1|p-1}
\sum_{\substack{\Delta\subset\overline{\Delta_{\lambda}},\\|\gamma_{\Delta}|\geq
1}} |\gamma_{\Delta}|^{p}.
\end{equation}
Quantity of cubes with $|\gamma_{\Delta}|=k$ is not more than
$\frac{|\gamma|}{k}$. Due to this, the following estimate holds:
\begin{align}\label{E:inequal2}
&\sum_{\substack{\Delta\in \overline{\Delta_{\lambda}},
\\|\gamma_{\Delta}|\geq1}}|\gamma_{\Delta}|^{p}=
\sum_{\substack{\Delta\in \overline{\Delta_{\lambda}},
\\|\gamma_{\Delta}|\geq p}}|\gamma_{\Delta}|^{p}+\sum_{k=1}^{p-1}
\sum_{\substack{\Delta\in \overline{\Delta_{\lambda}},
\\|\gamma_{\Delta}|=k}}|\gamma_{\Delta}|^{p}\leq\\
&\leq \sum_{\substack{\Delta\in \overline{\Delta_{\lambda}},
\\|\gamma_{\Delta}|\geq
p}}|\gamma_{\Delta}|^{p}+\sum_{k=1}^{p-1}k^{p-1}|\gamma|.\notag
\end{align}
Using \eqref{E:est7} and \eqref{E:inequal2}, we obtain the final
estimate of $U^{(p)}(\gamma)$:
\begin{equation}\label{E:final}
U^{(p)}(\gamma)\geq \sum_{\substack{\Delta\in
\overline{\Delta_{\lambda}},\\|\gamma_{\Delta}|\geq
p}}|\gamma_{\Delta}|^{p}\left(\frac {V_{p}^{p}(\Delta)}{p^p}-
pI_{p}^{1|p-1} \right)- pI_{p}^{1|p-1}\sum_{k=1}^{p-1}k
^{p-1}|\gamma|.
\end{equation}
Let us take into account the following obvious estimate:
\begin{equation}\label{E:ind_const}
\sum_{p>2} p\,I_{p}^{1|p-1}
\sum_{k=1}^{p-1}k^{p-1}<\sum_{p>2}p^{p+1}\,I_{p}^{1|p-1}.
\end{equation}
 The
condition of stability (superstability, strong superstability)
\eqref{E:main} - \eqref{E:main_addit} follows directly from the
last estimates \eqref{E:final}, \eqref{E:ind_const}  with
\begin{equation}\label{E:B}
 B=B_{2}+\sum_{p>2}p^{p+1}\,I_{p}^{1|p-1},
\end{equation}
where $B_{2}$ is taken from the condition of superstability(strong
superstability)of 2-body part of interaction.
\section{\bf Example of many-body interaction}
First consider one-dimensional case ($d=1$).
\begin{example}\label{E:examp1}
Let $V$ be a  many-body interaction, specified by a family of
$\textnormal{p}$-body potentials\,
$V_{p}:\left(\mathbb{R}^{d}\right)^{p}\rightarrow\mathbb{R},\,
p\geq2$:
\begin{align}\label{E:examp2}
&V_{p}(x_{1},\ldots,x_{p})=\frac{A_{p}}{\left(\underset{1\leq
i<j\leq
p}{\sum}|x_{i}-x_{j}|\right)^{m(p)}}-\frac{B_{p}}{\left(\underset{1\leq
i<j\leq p}{\sum}|x_{i}-x_{j}|\right)^{n(p)}},\\
&A_{p}>0, B_{p}>0;\; m(p)>n(p),\; n(p)>p-1. \notag
\end{align}
\end{example}
Prove that such a family of potentials satisfies assumptions {\bf
A1-A4}   and write down the conditions on $A_{p}, B_{p}$, that
ensure superstability of interaction.

Verification of assumptions {\bf A1-A3}\;is obvious. Let us analyze
the last assumption {\bf A4} for the family of p-body potentials
\eqref{E:examp2}. Denote the following area, which will be used in
our future estimates:
\begin{equation}\label{E:area1}
Q_{I_{p}}(x_{1})=\left\{ \{x_{2},\ldots,x_{p}\}\in
\mathbb{R}^{d(p-1)}\,|\,\sum_{1\leq i<j\leq p}|x_{i}-x_{j}|\geq
\left(\frac{A_{p}}{B_{p}}\right)^{\frac{1}{m(p)-n(p)}}\right\}
\end{equation}
 Using \eqref{E:area1} we can write
\begin{align}\label{E:examp3}
&I_{p}=\int_{\left(\mathbb{R}^{d}\right)^{p-1} } \; \left|
V_{p}^{-} \,\left(x_{1},\ldots,x_{p}\right )
\right|\;dx_{2}\cdot\ldots\cdot
dx_{p}=  \\
&=\int_{Q_{I_{p}}(x_{1}) }
\left(\frac{B_{p}}{\biggl(\underset{1\leq i<j\leq
p}{\sum}|x_{i}-x_{j}|\biggl)^{n(p)}}-\frac{A_{p}}{\biggl(\underset{1\leq
i<j\leq p}{\sum}|x_{i}-x_{j}|\biggl)^{m(p)}}\right)
\;dx_{2}\cdot\ldots\cdot dx_{p}.\notag
\end{align}
Consider the following estimates of sum $\underset{1\leq i<j\leq
p}{\sum}|x_{i}-x_{j}|$ in \eqref{E:examp2}. The minimum
of\;$\underset{1\leq i<j\leq p}{\sum}|x_{i}-x_{j}|$\; is reached, if
$p-2$ particles coincide. The maximum of\;$\underset{1\leq i<j\leq
p}{\sum}|x_{i}-x_{j}|$\; is reached, if $\left[\frac{p}{2} \right]$
particles are situated at one point and the rest of them are
situated at another one:
\begin{align}
&\underset{1\leq i<j\leq p}{\sum}|x_{i}-x_{j}|\geq
(p-1)\underset{1\leq i<j\leq p}{\textnormal{max}}|x_{i}-x_{j}|; \label{E:estmin}\\
&\underset{1\leq i<j\leq p}{\sum}|x_{i}-x_{j}|\leq
\left(p-\left[\frac{p}{2}\right]\right)\left[\frac{p}{2}\right]\underset{1\leq
i<j\leq p}{\textnormal{max}}|x_{i}-x_{j}|. \label{E:estmax}
\end{align}
\begin{remark}In d-dimensional case the estimate \eqref{E:estmin}
is also true, but the second one \eqref{E:estmax} requires the
following modification:
\begin{equation*}
\underset{1\leq i<j\leq p}{\sum}|x_{i}-x_{j}|\leq
\frac{p(p-1)}{2}\underset{1\leq i<j\leq
p}{\textnormal{max}}|x_{i}-x_{j}|.
\end{equation*}
\end{remark}
 Taking into account
\eqref{E:estmin},\,\eqref{E:estmax} we can rewrite
\eqref{E:examp3} in the following form:
\begin{align}\label{E:examp4}
&I_{p}\leq \int_{Q_{I_{p}}^{'}(x_{1}) } \;\left(
\frac{B_{p}^{'}}{\left(\underset{1\leq i<j\leq
p}{\textnormal{max}}|x_{i}-x_{j}|\right)^{n(p)}}-\frac{A_{p}^{'}}{\left(\underset{1\leq
i<j\leq p}{\textnormal{max}}|x_{i}-x_{j}|\right)^{m(p)}} \right)
\;dx_{2}\cdot\ldots\cdot dx_{p},
\end{align}
where
$A_{p}^{'}=\frac{A_{p}}{\left(\left(p-\left[\frac{p}{2}\right]\right)\left[\frac{p}{2}\right]\right)^{m(p)}},\quad
B_{p}^{'}=\frac{B_{p}}{(p-1)^{n(p)}}$,
\begin{equation}\label{E:area2}
Q_{I_{p}}^{'}(x_{1})=\left\{ \{x_{2},\ldots,x_{p}\}\in
\mathbb{R}^{d(p-1)}\,|\,\underset{1\leq i<j\leq
p}{\textnormal{max}}|x_{i}-x_{j}|\geq
\left(\frac{A_{p}^{'}}{B_{p}^{'}}\right)^{\frac{1}{m(p)-n(p)}}\right\}
\end{equation}
For definiteness we will assume, that: $x_{1}=0$. There are two
types of configurations:
\begin{align}
&1)
\,\textnormal{diam}(\{x_{1},\ldots,x_{p}\})=\textnormal{dist}(x_{i};x_{j}),\,
 \textnormal{for some} x_{i},x_{j}\, \textnormal{and} x_{i}<x_{1}<x_{j}; \notag \\
&2)\,\textnormal{diam}(\{x_{1},\ldots,x_{p}\})=\textnormal{dist}(x_{1};x_{j}),\,
\textnormal{for some} x_{j} \notag
\end{align}
In accordance with these 2 cases we can rewrite \eqref{E:examp4}
in the following form:
\begin{align}
&I_{p}\leq
A_{p-1}^{2}\int_{-\infty}^{-\bigl(\frac{A_{p}^{'}}{B_{p}^{'}}\bigl)^{\frac{1}{m(p)-n(p)}}}\,dx_{2}\cdot
\int_{0}^{+\infty} \left(
\frac{B_{p}^{'}}{(x_{p}-x_{2})^{n(p)}}-\frac{A_{p}^{'}}{(x_{p}-x_{2})^{m(p)}}\right)\,dx_{p}\times
\notag
\\
&\times \int_{x_{2}}^{x_{p}}dx_{3}\cdots
\int_{x_{2}}^{x_{p}}\,dx_{p-1}+  \notag \\
&+A_{p-1}^{2}\int_{-\bigl(\frac{A_{p}^{'}}{B_{p}^{'}}\bigl)^{\frac{1}{m(p)-n(p)}}}^{0}\,dx_{2}\cdot
\int_{x_{2}+\bigl(\frac{A_{p}^{'}}{B_{p}^{'}}\bigl)^{\frac{1}{m(p)-n(p)}}}^{+\infty}
\left(
\frac{B_{p}^{'}}{(x_{p}-x_{2})^{n(p)}}-\frac{A_{p}^{'}}{(x_{p}-x_{2})^{m(p)}}\right)\,dx_{p}\times \notag\\
&\times \int_{x_{2}}^{x_{p}}dx_{3}\cdots
\int_{x_{2}}^{x_{p}}\,dx_{p-1}+  \notag \\
&+2(p-1)\int_{\bigl(\frac{A_{p}^{'}}{B_{p}^{'}}\bigl)^{\frac{1}{m(p)-n(p)}}}^{+\infty}
\left(
\frac{B_{p}^{'}}{x_{p}^{n(p)}}-\frac{A_{p}^{'}}{x_{p}^{m(p)}}\right)\,dx_{p}\,
\int_{0}^{x_{p}}dx_{2}\cdots
\int_{0}^{x_{p}}\,dx_{p-1}\label{E:examp5}.
\end{align}
The first and the second integrals  in \eqref{E:examp5}
 refer to the case 1). In
\eqref{E:examp5}\,$A_{p-1}^{2}$ is a number of all possible pairs
$(x_{i},x_{j}),\; 1< i<j\leq p$ with respect to their order:\\
$A_{n}^{k}=\frac{n!}{(n-k)!}$. The third integral in
 \eqref{E:examp5}\,refers to the case 2). In \eqref{E:examp5}\,
$2(p-1)$ is a number of $x_{i},\;1< i\leq p $ with respect to its
right or left position from the origin. Under the condition \,
$m(p)>n(p),\; n(p)>p-1$\, all integrals \eqref{E:examp5}\,converge
and finally:
\begin{align}\label{E:examp8}
&I_{p}\leq\left(\frac{A_{p}^{'}}{B_{p}^{'}}\right)^{\frac{p-1}{m(p)-n(p)}}\Biggl(
A_{p-1}^{2}\Biggl( \frac{A_{p}^{'}}{(p-2-m(p))
\left(\frac{A_{p}^{'}}{B_{p}^{'}}\right)^{\frac{m(p)}{m(p)-n(p)}}}
-\frac{B_{p}^{'}}{(p-2-n(p))\left(\frac{A_{p}^{'}}{B_{p}^{'}}\right)^{\frac{n(p)}{m(p)-n(p)}}}+\notag\\
&+\frac{A_{p}^{'}(-1)^{p-2-m(p)}}{(p-2-m(p))(p-1-m(p))}-\frac{B_{p}^{'}(-1)^{p-2-n(p)}}{(p-2-n(p))(p-1-n(p))}\Biggl)+\notag\\
&+2(p-1)\Biggl(
\frac{A_{p}^{'}}{(p-1-m(p))\left(\frac{A_{p}^{'}}{B_{p}^{'}}\right)^{\frac{m(p)}{m(p)-n(p)}}}-\frac{B_{p}^{'}}{(p-1-n(p))
\left(\frac{A_{p}^{'}}{B_{p}^{'}}\right)^{\frac{n(p)}{m(p)-n(p)}}}\Biggl)\Biggl).
\end{align}
Consequently, the assumption {\bf A4}\,holds. We obtain  from
\eqref{E:main_addit}, \eqref{E:examp8} the condition on $A_{p},
B_{p}$, that ensures  supersable interaction:
\begin{equation}\label{E:A_p_B_p_cond}
  \sum_{p\geq2}p^{p+1}\,I_{p}<+\infty,
\end{equation}
so we can put for example:
\begin{equation}\label{E:examp10}
n(p)=p,\, m(p)=p+1,\,
A_{p}=\left(\left(p-\left[\frac{p}{2}\right]\right)\left[\frac{p}{2}\right]\right)^{p+1},\,
B_{p}<\frac{p^{\frac{p-4}{2}-\varepsilon}}{2^{\frac{2p-1}{2}}},\,
\varepsilon>0.
\end{equation}
Let us find an estimate of\, $U^{(3)}(\gamma)$. For simplicity:\,
$A_{3}=B_{3}=1;\,m(3)=12,\,n(3)=6$. Then:\,$A_{3}^{'}=1/2^{12},\,
B_{3}^{'}=1/2^{6}$.\, Using \eqref{E:examp8},\, we
obtain:\,$I_{3}\leq \frac{477}{20480}$. Taking into account
Remark~\ref{R:converg}, we conclude, that:
$I_{3}^{1|2}(\Delta)\lambda^2\rightarrow \frac{477}{20480},\,
\lambda\rightarrow 0$. Now estimate\, $V_{3}^{3}(\Delta)$,\, where\,
$\Delta\in\overline{\Delta_{\lambda}},\;V_{3}^{3}(\Delta)\geq 0,
\textnormal{for any} \{x_{1},\cdots,x_{p}\}\subset{}\Delta $.
 It follows from \eqref{E:estmin},\,\eqref{E:estmax},\, that:
\begin{equation}\label{E:examp9}V_{p}^{p}(\Delta)\geq \frac{A_{p}^{'}}{\left(\underset{1\leq
i<j\leq
p}{\textnormal{max}}|x_{i}-x_{j}|\right)^{m(p)}}-\frac{B_{p}^{'}}{\left(\underset{1\leq
i<j\leq p}{\textnormal{max}}|x_{i}-x_{j}|\right)^{n(p)}}.
\end{equation}
Function in the right part of \eqref{E:examp9} achieves its minimum
in the cubic area, if:\\$\underset{1\leq i<j\leq
p}{\textnormal{max}}|x_{i}-x_{j}|=\lambda$,\, where $\lambda$ is a
rib of $\Delta$. Using \eqref{E:examp9}, we
obtain:\,$V_{3}^{3}(\Delta)\geq\frac{1}{4096\lambda^{12}}-\frac{1}{64\lambda^{6}}$.
We have finally from \eqref{E:final}:
\[U^{(3)}(\gamma)\geq \sum_{\substack{\Delta\in
\overline{\Delta_{\lambda}},\\|\gamma_{\Delta}|\geq
3}}|\gamma_{\Delta}|^{3}\left(\frac {1}{110592\lambda^{12}}-\frac
{1}{1728\lambda^{6}}- \frac{1431}{20480\lambda^{2}}\right)-
\frac{1431}{4096 \lambda^{2}}|\gamma|.\]
 If \,$\lambda\,\leq 0.29874$, then condition \eqref{E:main} of
 Theorem \ref{T:main} holds.

Now, consider the general case $d>1$ with $B'_{p}>0$ and
$n(p)>(p-1)d$. It is clear that
\[\left|
V_{p}^{-} \,\left(x_{1},\ldots,x_{p}\right ) \right|\leq
\frac{B'_{p}}{\left(\underset{1\leq i<j\leq
p}{\textnormal{max}}|x_{i}-x_{j}|\right)^{n(p)}},\]
Prove that it satisfies assumption {\bf A4}.\\
{\it Proof.} Consider such a ball $B(0;R_{0})$ with center in the
origin and a radius $R_{0}$ that
$V_{p}\left(x_{1},\ldots,x_{p}\right)\geq 0\, \text{for any }\,
p\geq 2$ and put $x_{1}=0$. As in the case $d=1 $ consider 2 cases:
\begin{align}
&1)\,\textnormal{diam}(\{x_{1},\ldots,x_{p}\})=\textnormal{dist}(x_{i};x_{j}),\notag \\
&0 \in B\left(
\frac{x_{i}+x_{j}}{2};\,\frac{|x_{j}-x_{i}|}{2}\right), 1<i\leq p,
1<j\leq p; \notag\\
&2)\,\,\textnormal{diam}(\{x_{1},\ldots,x_{p}\})=\textnormal{dist}(0;x_{j}),
\,1<j\leq p.\notag
\end{align}
In accordance with these 2 cases one can write the following
estimate:
\begin{align}
&\int_{\left(\mathbb{R}^{d}\right)^{p-1}} \; \left| V_{p}^{-}
\,\left(0, x_{2},\ldots,x_{j}\right )
\right|\;dx_{2}\cdots dx_{p}\leq I_{p}^{(1)}+I_{p}^{(2)}; \notag\\
&I_{p}^{(1)}\leq B'_{p}\cdot
C_{p-1}^{2}\underset{\substack{\frac{|x_{2}+x_{p}|}{2}\leq\frac{|x_{p}-x_{2}|}{2},\\
|x_{p}-x_{2}|>2R_{0}
}}{\int}\,\frac{dx_{2}\cdot\,dx_{p}}{|x_{p}-x_{2}|^{n(p)}}\times
\label{E:generalize1}
\\
&\times
\underset{\left|\frac{x_{p}+x_{2}}{2}-x_{3}\right|\leq\frac{|x_{p}-x_{2}|}{2}}{\int}dx_{3}\cdots
\underset{\left|\frac{x_{p}+x_{2}}{2}-x_{p-1}\right|\leq\frac{|x_{p}-x_{2}|}{2}}{\int}\,dx_{p-1}, \notag\\
&I_{p}^{(2)}\leq B'_{p}(p-1)\,
\underset{|x_{p}|>2R_{0}}{\int}\,\frac{dx_{p}}{|x_{p}|^{n(p)}}\cdot
\underset{\left|\frac{x_{p}}{2}-x_{2}
\right|\leq\frac{|x_{p}|}{2}}{\int}\, dx_{2} \cdots
\underset{\left|\frac{x_{p}}{2}-x_{p-1}
\right|\leq\frac{|x_{p}|}{2}}{\int}\, dx_{p-1}\label{E:generalize2}.
\end{align}
The first integral $I_{p}^{(1)}$ and the second $I_{p}^{(2)}$
refer to cases 1) and 2) respectively. In
\eqref{E:generalize1}\\$C_{p-1}^{2}$ is a  quantity of all
possible pairs $\{x_{i},x_{j}\},\; 1< i<j\leq p$ without respect
to their order. In \eqref{E:generalize2} $p-1$ is a quantity of
$x_{j},\;1< j\leq p $. The case of d-dimensional space differs
from 1-dimensional in the way, that the order of the "remotest"
variables \,$x_{i},x_{j}$ is neglected. Let's take into account
that a volume of d-dimensional ball $B(a, R )$ is:
\begin{equation}\label{E:volume}
\int_{|x-a|\leq
R}\,dx=\frac{2\pi^{\frac{d}{2}}R^{d}}{d\,\Gamma\left( \frac{d}{2}
\right)}.
\end{equation}
Using \eqref{E:volume} one can rewrite\, \eqref{E:generalize1},\,
\eqref{E:generalize2}\, in the following form:
\begin{align}
&I_{p}^{(1)}\leq B'_{p}\,
C_{p-1}^{2}\,\left(\frac{\pi^{\frac{d}{2}}}{2^{d-1}d\,\Gamma\left(\frac{d}{2}\right)}\right)^{p-3}\underset{\substack{\frac{|x_{2}+x_{p}|}{2}\leq\frac{|x_{p}-x_{2}|}{2},\\
|x_{p}-x_{2}|>2R_{0}
}}{\int}\,\frac{dx_{2}\cdot\,dx_{p}}{|x_{p}-x_{2}|^{n(p)-(p-3)d}},
\label{E:gerneralize3}\\
&I_{p}^{(2)}\leq
B'_{p}\,(p-1)\,\left(\frac{\pi^{\frac{d}{2}}}{2^{d-1}d\,\Gamma\left(\frac{d}{2}\right)}\right)^{p-2}
\underset{|x_{p}|>2R_{0}}{\int}\,\frac{dx_{p}}{|x_{p}|^{n(p)-(p-2)d}}.
\label{E:gerneralize4}
\end{align}
In \eqref{E:gerneralize3} we do the following substitution of
variables: $\{x_{2};x_{p}\}\rightarrow \{x_{2};t\},\,
t=x_{p}-x_{2}$.\\
 We obtain:
\begin{align}
&I_{p}^{(1)}\leq B'_{p}\,
C_{p-1}^{2}\,\left(\frac{\pi^{\frac{d}{2}}}{2^{d-1}d\,\Gamma\left(\frac{d}{2}\right)}\right)^{p-3}
\underset{|t|>2R_{0}}{\int}\frac{dt}{|t|^{n(p)-(p-3)d}}\,
\underset{|x_{2}+\frac{t}{2}|\leq \frac{|t|}{2}}{\int}\,dx_{2}= \notag \\
 &=B'_{p}\,
C_{p-1}^{2}\,\left(\frac{\pi^{\frac{d}{2}}}{2^{d-1}d\,\Gamma\left(\frac{d}{2}\right)}\right)^{p-2}
\underset{|t|>2R_{0}}{\int}\frac{dt}{|t|^{n(p)-(p-2)d}}\label{E:generalize5}
\end{align}
Using generalized spherical coordinates, we deduce from\,
\eqref{E:gerneralize4}\, and\, \eqref{E:generalize5}\, that:
\begin{align}
&I_{p}^{(1)}\leq B'_{p}\,
C_{p-1}^{2}2^{d}d\,\left(\frac{\pi^{\frac{d}{2}}}{2^{d-1}d\,\Gamma\left(\frac{d}{2}\right)}\right)^{p-1}
\int_{2R_{0}}^{+\infty}\frac{dr}{r^{n(p)+(1-p)d+1}}\label{E:generalize6},\\
&I_{p}^{(2)}\leq B'_{p}\,
(p-1)2^{d}d\,\left(\frac{\pi^{\frac{d}{2}}}{2^{d-1}d
\,\Gamma\left(\frac{d}{2}\right)}\right)^{p-1}
\int_{2R_{0}}^{+\infty}\frac{dr}{r^{n(p)+(1-p)d+1}}\label{E:generalize7}.
\end{align}
If $n(p)+(1-p)d>0$ integrals \eqref{E:generalize6}\, and
\eqref{E:generalize7} converge and finally:
\begin{align}
&\int_{\left(\mathbb{R}^{d}\right)^{p-1}} \; \left| V_{p}^{-}
\,\left(0, x_{2},\ldots,x_{p}\right ) \right|\;dx_{2}\cdots
dx_{p}\leq \frac{B'_{p}\, d\, \left( p-1+\, C_{p-1}^{2} \right)}
{2^{n(p)-pd}(n(p)+(1-p)d)R_{0}^{n(p)+(1-p)d} }\times \notag \\
&\times \left(\frac{\pi^{\frac{d}{2}}}{2^{d-1}d
\,\Gamma\left(\frac{d}{2}\right)}\right)^{p-1}
\label{E:generalize8}.
\end{align}
In this case condition on $B'_{p}$, which ensures superstability,
can be easily obtained in a similar way as in the example
\eqref{E:examp2}.
$$
\mspace{675mu}\blacksquare
$$
\section{\bf Appendix}
For  arbitrary $a_{1},\dots,a_{j}\in
\mathbb{R}_{+};m_{1},\dots,m_{j}\in \mathbb{N}$ the following
inequality holds:
\begin{equation}\label{A:eq1}
\prod_{i=1}^{j}a_{i}^{m_{i}}\leq\frac{1}{m_{1}+\cdots+m_{j}}\sum_{i=1}^{j}m_{i}\,a_{i}^{m_{1}+\cdots+m_{j}}.
\end{equation}
\emph{Proof.}
Let's prove \eqref{A:eq1} in the  case of $j=2$.\\
Let $j=2, m_{2}=1$  in \eqref{A:eq1}. Then the following is true:
\begin{equation}\label{A:l1}
a_{1}^{m_{1}}a_{2}\leq
\frac{m_{1}}{m_{1}+1}a_{1}^{m_{1}+1}+\frac{1}{m_{1}+1}a_{2}^{m_{1}+1}.
\end{equation}
We will prove this fact by induction. If $m_{1}=1$, then
inequality \eqref{A:l1} transforms into trivial: $a_{1}a_{2}\leq
\frac{a_{1}^{2}}{2}+\frac{a_{2}^{2}}{2}$. We suppose, that
\eqref{A:l1} is true for $m_{1}-1$. Taking into account this
statement, the following range of estimates holds:
\begin{equation}\label{A:l2}
a_{1}^{m_{1}}a_{2}=a_{1}^{m_{1}-1}(a_{1}a_{2})\leq
\frac{a_{1}^{m_{1}+1}}{2}+\frac{a_{1}^{m_{1}-1}a_{2}^{2}}{2}\leq
\frac{a_{1}^{m_{1}+1}}{2}+\frac{a_{2}}{2}\left[
\frac{m_{1}-1}{m_{1}}a_{1}^{m_{1}}+\frac{1}{m_{1}}a_{2}^{m_{1}}\right].
\end{equation}
From \eqref{A:l2} we obtain:
\begin{equation}\label{A:l3}
\frac{m_{1}+1}{2m_{1}}a_{1}^{m_{1}}a_{2}\leq\frac{1}{2}a_{1}^{m_{1}+1}+\frac{1}{2m_{1}}a_{2}^{m_{1}+1}.
\end{equation}
The estimate \eqref{A:l1} follows directly from \eqref{A:l3}. We
deduce from \eqref{A:l1}:
\begin{align}
&a_{1}^{m_{1}}a_{2}^{m_{2}}=a_{2}^{m_{2}-1}(a_{1}^{m_{1}}a_{2})\leq\frac{m_{1}}{m_{1}+1}a_{1}^{m_{1}+1}a_{2}^{m_{2}-1}+
\frac{1}{m_{1}+1}a_{2}^{m_{1}+m_{2}}\leq \label{A:eq2}\\
&\leq
a_{2}^{m_{2}-2}\frac{m_{1}}{m_{1}+1}\left[\frac{m_{1}+1}{m_{1}+2}a_{1}^{m_{1}+2}+\frac{1}{m_{1}+2}a_{2}^{m_{1}+
2}\right]+\frac{1}{m_{1}+1}a_{2}^{m_{1}+m_{2}}= \notag \\
&=\frac{m_{1}}{m_{1}+2}a_{1}^{m_{1}+2}a_{2}^{m_{2}-2}+\frac{2}{m_{1}+2}a_{2}^{m_{1}+
m_{2}}\leq\ldots\leq
\frac{m_{1}}{m_{1}+m_{2}}a_{1}^{m_{1}+m_{2}}+\frac{m_{2}}{m_{1}+m_{2}}a_{2}^{m_{1}+
m_{2}}\notag.
\end{align}
The estimate \eqref{A:eq1} is proven for $j=2$. Let \eqref{A:eq1}
be true for $j-1$. Using this, we obtain the following range of
estimates:
\begin{align}
&\prod_{i=1}^{j}a_{i}^{m_{i}}\leq
\frac{1}{m_{1}+\cdots+m_{j-1}}\sum_{i=1}^{j-1}m_{i}\,a_{i}^{m_{1}+\cdots+m_{j-1}}\;a_{j}^{m_{j}}\leq
\label{A:eq3}
\\
&\leq
\frac{1}{m_{1}+\cdots+m_{j-1}}\sum_{i=1}^{j-1}\left[\frac{m_{i}(m_{1}+\cdots+m_{j-1})}{m_{1}+\cdots+m_{j}}a_{i}^{m_{1}+\cdots+m_{j}}+
\frac{m_{i}m_{j}}{m_{1}+\cdots+m_{j}}a_{j}^{m_{1}+\cdots+m_{j}}\right]\notag.
\end{align}
One can deduce from the last estimate in \eqref{A:eq3}:
\begin{align}
&\prod_{i=1}^{j}a_{i}^{m_{i}}\leq
\frac{1}{m_{1}+\cdots+m_{j}}\sum_{i=1}^{j-1}m_{i}\,a_{i}^{m_{1}+\cdots+m_{j}}+\label{A:eq4}\\
&+\frac{m_{j}}{(m_{1}+\cdots+m_{j-1})(m_{1}+\cdots+m_{j})}\,a_{j}^{m_{1}+\cdots+m_{j}}\sum_{i=1}^{j-1}m_{i}\notag.
\end{align}
The estimate \eqref{A:eq1} is a consequence of \eqref{A:eq4}. The
end of the proof.
$$
\mspace{675mu}\blacksquare
$$

\textbf{Acknowledgments.} The author is grateful to Prof.
O.~L.~Rebenko for the setting of this problem  and stimulating
discussions concerning the subject of this paper.

\vskip-1cm
\renewcommand{\refname}{}


\vskip-1cm

\begin{thebibliography}{GKKS01}

\bibitem{BePe2002}
V.~Belitsky, E.~A.~Pechersky,
\newblock{Uniqueness of Gibbs state for nonideal gas in $\mathbb{R}^{d}$.  The case of multibody
interaction},
\newblock{\em J.~Stat.~Phys.}, {\bf 106}, 931-955 (2002).

\bibitem{Ge82}W.~Greenberg,
\newblock{Thermodynamic states of classical systems},
\newblock{\textit{Commun.Math. Phys.}}, {\bf 22}, 259-268(1971).

\bibitem{Mo76}H. Moraal,
\newblock{The Kirkwood-Salzburg equation and the virial expansion for many-body
potentials},
\newblock{\em Phys. Lett.}, {\bf 59A}, 9-10(1976).

\bibitem{PR07}
S.~N.~Petrenko, A.~L.~Rebenko,  Superstable criterion and
superstable bounds for infinite range interaction I: two-body
potentials,  \textit{Meth. Funct. Anal. and Topology}, {\bf 13},
50--61(2007).

\bibitem{PS2000}
A.~Procacci, B.~Scoppola,
\newblock {The gas phase of continupous systems of hard spheres interacting via n-body
potential},
\newblock {\em Commun.~Math.~Phys.}, {\bf 211}, 487-496(2000).

\bibitem{ReK04}
 O.~V.~Kutoviy, A.~L.~Rebenko,
\newblock { Existence of Gibbs state for continuous gas with many-body interection,}
\newblock {\em J.~Math.~Phys. }, {\bf 45(4)},  1593-1605 (2004).

\bibitem{RS97}
A.~L.~Rebenko, G.~V.~Shchepan'uk,
\newblock {The convergence of cluster expansions for continuous
systems with many-body interactions},
\newblock{\em J.~Stat.~Phys.}, {\bf 88}, 665-689 (1997).

\bibitem{RT07}
A.~L.~Rebenko, M.V. Tertychnyi,  On the Superstability and Strong
Superstability of 2-Body Interaction Potentials,
\newblock {\em \textit{Meth. Funct. Anal. and Topology}}, {\bf ?}
?-?(2008).

\bibitem{R70}
D.~Ruelle,
\newblock { Superstable interactions in classical statistical
mechanics,}
\newblock {\em Commun.~Math.~Phys.}, {\bf 18},  127-159 (1970).

\bibitem{Sk06}
W.~I.~Skrypnik,
\newblock {On Gibbs quantum and classical particle systems with three-body
forces},
\newblock{\em  Ukrainian Mathematical Journal}, {\bf 58(7)},
976-996 (2006).

\bibitem{Za81}
V.~A.~Zagrebnov,
\newblock {On the solution of correlation equations for classical continuous systems},
\newblock{\em  Physica A}, {\bf 109},
403-424 (1981).
 \end{thebibliography}
\end{document}